\documentclass[journal]{IEEEtran}

\author {A.~Grigoryants, Russian-Armenian State University}
\title {On a Kronecker products sum distance bounds}

\newcommand {\B} [1] {\mathbf {#1}}
\newtheorem {theorem}	{Theorem}
\newtheorem {lemma}	{Lemma}
\newtheorem {example}	{Example}
\newtheorem {conc}	{Conclusion}

\begin{document}
\maketitle

\begin{keywords}
tensor product, fractal code, distance, bound, lower, upper, binary error correcting code
\end{keywords}

\abstract{ 
A binary linear error correcting codes represented by two code families 
Kronecker products sum are considered.
The dimension and distance  of new code is investigated.
Upper and lower bounds of distance are obtained.
Some examples are given.
It is shown that some classic constructions are the private cases of 
considered one. The subclass of codes with equal lower and 
upper distance bounds is allocated.}

\section{Introduction.}

This article is an introduction to projected circle of articles on 
researching of binary error correcting codes 
(hereinafter -- \emph{codes}) represented by two codes' families
Kronecker products sum.
This codes' family  is well known and considered, for example, in
\cite{stif}, \cite{march}, \cite{licin}, \cite{boyar}.
Some authors call these codes ``Kronecker sums'' \cite{licin}.
We suggest the term \emph{fractal codes} for this codes.
So the code on length $n = n_1\cdot n_2$
is built from two code families on lengths $n_1$ and $n_2$ as
sum of tensor products of corresponding codes.
The dimension and distance  of new code are investigated.
Upper and lower bounds of distance are obtained.

The generalized cascade codes are the particular case of fractal codes.
Most of our results are proved for so called acyclic codes' families.
If one of codes' families is acyclic and the other is embedded, we state 
the subclass of fractal codes with equal upper and lower distance 
bounds.

\section{Basic definitions}

We say $(n,k,d)$-code meaning a linear binary code 
of length $n$, dimension $k$ and distance $d$, i.e.
$k$-dimensional subspaces of linear space
$\B{F}_{2}^{n}$ over the field $\B{F}_{2}=\{0;1\}$. 

We use standard terminology of algebraic coding theory
(we follow~\cite{mac}) and we use basic concepts of linear and tensor algebra with 
no comments. Let
$\B{C}=\{C_{i}\}_{i=\overline{1;s}}$ and
$\B{D}=\{D_{i}\}_{i=\overline{1;s}} $
be two binary codes' families of length
$n$ and $n'$ respectively. 
Let us consider the code: 
\begin{equation}\label{f1}
	\B{C\otimes D} = C_{1}\otimes D_{1}+\dots+C_{s}\otimes D_{s},
\end{equation}
So (\ref{f1}) is the sum of tensor products of corresponding codes from 
the families $\B{C}$ and $\B{D}$. 
Expression (\ref{f1}) gives the code of length 
$nn'$. We'll call the codes represented by (\ref{f1}) for some two 
families $\B{C}$ and $\B{D}$ \emph{fractal codes}.  
In this paper the basic parameters of construction (\ref{f1}) will be
researched.

Let $S = \{1;\dots ;s\}$ be the set of first $s$ natural numbers
and 
$ \alpha =\{i_{1};\dots ;i_{r} \} \subset S $ be an arbitrary subset
in $S$. Note $|\alpha | = r$.
For an arbitrary codes family
$\{C_{i}\}_{i=\overline{1;s}}$ we'll consider following codes:
\begin {eqnarray} \label{f2}
	C^{\alpha}=\sum_{i \in \alpha}{C_{i}} \nonumber\\
	C_{\alpha}=\bigcap_{i \in \alpha}{C_{i}}
\end {eqnarray}
Thus $C^{\alpha}$ и $C_{\alpha}$ are respectively the sum and intersection 
of linear spaces $C_i$ where $i$ runs over the $\alpha$.

We also write 
$C_{12}$ for $C_1\cap C_2$ etc. Denote 
$(n^{\alpha}, k^{\alpha}, d^{\alpha})$  the parameters of codes $C^\alpha$ and
$(n_{\alpha},k_{\alpha}, d_{\alpha})$ the parameters of codes $C_\alpha$.
Let 
$(n'^{\alpha}, k'^{\alpha}, d'^{\alpha})$ and
$(n'_{\alpha},k'_{\alpha},d'_{\alpha})$ be the respective codes parameters
for $\B{D}$ family.

We say a vectors family $\B{e}=\{e_i\} \subset \cup{\B{C}}$ to be
\emph{basis} of $\B{C}$, if:
\begin {enumerate}
  \item { $C_\alpha\cap\,\B{e} $ generates $C_\alpha$ }
  \item {$\B{e}$ is minimal (by inclusion) family with property 1}
\end {enumerate}

It's easy to see that arbitrary family of subspaces has a basis but basis can
be a linear dependent family. The detailed research of family basis properties is out
of this paper boundaries.

The family of subspaces with linear independent basis is called \emph{acyclic family}.

Let us denote $\alpha_b$ the maximal by power multi-index with property
$b \in C_{\alpha_b}$. It's obviously defined uniquely.
For arbitrary vectors family $\B{b}=\{b_i\}$
we denote $\Psi(\B{b})=\{\alpha_{b_i} \;|\; b_i \in \B{b}\}$.
Let now $\Psi$ is an arbitrary family of multi-indexes.
Choosing one element from each multi-index in
$\Psi$ we get some multi-index.
The set of all such multi-indexes are denoted
$\Psi^*$.

The main result of this paper is represented in the following theorem:
\begin {theorem} \label {theorem:A}
	Let $\B{C}$ and $\B{D}$ be acyclic codes' families.
	Then the dimension of code (\ref{f1}) can be calculated by formula:
\begin {equation} \label{f3}
	\kappa=\sum_{\alpha}{(-1)^{|\alpha|+1}k_{\alpha}k'_{\alpha}}
\end {equation}

Upper distance bound of code (\ref{f1}):
\begin {equation} \label{f4}
\delta 	\leq \min_{\alpha}{(d_{\alpha}d'^{\alpha},
d^{\beta}d'_{\beta})}
\end {equation}
where minimum is calculated by all
$\alpha, \beta$ with nonzero subspaces $C_{\alpha},D_{\beta}$.

Let $\B{e}=\{e_i\}$ be a basis in $\B{C}$ and
$\B{g}=\{g_j\}$ be a basis in  $\B{D}$. 
Then lower distance bound for code (\ref{f1}):
\begin {eqnarray} \label{f4g}
	\delta 	\geq
	\max
	\left(
		\min_{\Psi_0 \subset \Psi(\B{e})}
		{\left[
			(\max_{\alpha \in\Psi_0} {d'^\alpha})
			(\max_{\beta\in\Psi^*_0} {d^\beta})
		 \right]
		} , \right. \\ \nonumber \left .
		\min_{\Psi_0 \subset \Psi(\B{g})}
		{\left[
			(\max_{\alpha \in\Psi_0} {d^\alpha})
			(\max_{\beta\in\Psi^*_0} {d'^\beta})
		 \right]
		}
	  \right)
	,
\end {eqnarray}
where $\Psi_0$ is arbitrary nonempty subset in $\Psi(\B{e})$ or $\Psi(\B{g})$.
\end {theorem}

We call family $\B{C}$ \emph{embedded family}, if $C_i \subset C_{i+1}$ for all 
$i$. Embedded family is obviously acyclic.

The following theorem describes the subclass 
with coincided upper and lower distance bounds:
\begin {theorem} \label {theorem:B}
	If one of families $\B{C, D}$ is embedded and another is acyclic
	then upper (\ref {f4}) and lower (\ref {f4g}) distance bounds are coincided
	and therefore upper bound (\ref {f4}) reached.
\end {theorem}

\begin {conc} \label {conc:c1}
	If $\B{C}$ and $\B{D}$ are embedded, then the dimension and distance
	of code
	\begin {equation} \label {f6s}
	  E = C_1 \otimes D_s + C_2 \otimes D_{s-1} + \dots + C_s \otimes D_1
	\end {equation}
	can be calculated by formulas:
	\begin {eqnarray*}
	\kappa &=& k_1 k'_s + (k_2-k_1)k'_{s-1} + \dots + (k_s-k_{s-1})k'_1  \\
	\delta &=& \min_i (d_id_{s-i+1})
	\end {eqnarray*}
\end {conc}

\section {The theorems proof.}

We need some auxiliary lemmas to prove theorems.
The following simple approval often is useful.
\begin{lemma}\label{lemma:l10}
	Let $L,M$ be two linear spaces. Then for arbitrary vector
	$x$ from $L\otimes M$ there exists unique representation of type:
\begin {equation}\label{f4_5}
	x=\sum_i {e_i\otimes b_i}
\end {equation}
where $\{e_i\}$ is a basis in $L$ and $b_i$ are some vectors from $M$.
In particular $x=0$ then and only then all $b_i=0$.
\end{lemma}
\begin{proof}
Let $\{g_i\}$ be basis in $M$. Then $\{e_i\otimes g_j\}$ is
basis in $L\otimes M$ and arbitrary $x$ from $L\otimes M$ uniquely
is implemented as:
\begin {displaymath}
	x=\sum_{i,j}{a_{ij}(e_i\otimes g_j)}=\sum_i{e_i\otimes
	\sum_j{a_{ij}g_j}}=\sum_i{e_i\otimes b_i},
\end {displaymath}
where we denote $\sum_j{a_{ij}g_j}\in M$ as $b_i$.
\end{proof}

\begin{lemma}\label {lemma:l11}
	If family $\B{C}$ is acyclic, then for arbitrary vector
	$x \in \B{C \otimes D}$ there exists unique representation:
	\begin {equation} \label {f5}
		x = e_1 \otimes b_1 + \dots + e_r \otimes b_r \quad ,
	\end {equation}
	where $\B{e}=\{e_i\}$ is a basis in $\B{C}$ and
		$ b_i \in D^{\alpha_{e_i}} $.

\end {lemma}
\begin{proof}
Vector $x$ can be represented as
$x=x_1 + \dots + x_s$, where $x_i \in C_i \otimes D_i$,
because $x \in \B{C \otimes D}$. Every $x_i$
according to lemma \ref {lemma:l10} can be represented as :
\begin {displaymath}
	x_i = \sum_j {e^j_i \otimes b^j_i} \quad,
\end {displaymath}
where $e^j_i\in \B{e}\cap C_i$ and $b^j_i\in D_i$ for all $i,j$.
Grouping similar members (i.e. bearing out of branches same vectors
$e^j_i$), we get representation of type (\ref {f5}), where, obviously, $b_i\in
D^{\alpha_{e_i}} $. Uniquety implies from the following fact:
for arbitrary vector $y=\sum_i{a_i\otimes b_i}$ the linear independence
of $a_i$ vectors implies that $y=0\Leftrightarrow b_i=0$ for all $b_i$.
\end{proof}

\begin {lemma}\label{lemma:l20}
Let $L,M$ be two linear spaces. $L_1,\dots,L_s \subset L$ and
$M_1,\dots, M_s \subset M $  are linear subspaces in $L$ and $M$.
Then the equation take place:
	\begin{equation}\label{f5s}
		(L_1\otimes M_1)\cap\dots\cap (L_s\otimes M_s) =
		(L_1\cap\dots\cap L_s)\otimes (M_1\cap\dots\cap  M_s)
	\end{equation}
\end {lemma}
\begin{proof}
First let us consider the case $s=2$.
Note that if $L_{12}=0$ or $M_{12}=0$, then
\begin {displaymath}
	(L_1 \otimes M_1) \cap (L_2 \otimes M_2) = 0
\end {displaymath}

Indeed, let, for example, $L_{12}=0$. Then for arbitrary bases
$\{e^1_i\}$ in $L_1$ $\{e^2_j\}$ in $L_2$
and for any vector $x$ from intersection we can write:
\begin {displaymath}
	x = \sum_i{e^1_i\otimes b_i} =
	\sum_j{e^2_j\otimes b'_j}
\end {displaymath}
But according to lemma \ref{lemma:l10}, considering linear independence
of system $\{e^1_i,e^2_j\}$ we get $x=0$.

Tensor product is distributive above direct sum.
The following computation uses this fact.  
Denote $L'_i\; (M'_i)$ an arbitrary direct complement
$L_{12}\; (M_{12})$ in $L_i\; (M_i)$, where
$i=1,2$. We have:
\begin {displaymath}
\begin {array}{ccc}
	(L_1\otimes M_1)\cap (L_2\otimes M_2 ) =
	((L_{12}\oplus L'_1)\otimes (M_{12}\oplus M'_1))
	\cap \\ \cap ((L_{12}\oplus L'_2)\otimes (M_{12}\oplus M'_2)) =
	L_{12}\!\otimes\! M_{12}\oplus \\ \oplus (
	L_{12}\!\otimes\! M'_1\oplus L'_1\!\otimes\! M_1)\cap
	L_{12}\!\otimes\! M_{12}\oplus(
	L_{12}\!\otimes\! M'_2\oplus L'_2\!\otimes\! M_2)
\end {array}
\end {displaymath}
Let us denote:
$ 	A=L_{12}\!\otimes\! M_{12},\;
	B=L_{12}\!\otimes\! M'_1\oplus L'_1\!\otimes\! M_1,\;
	C=L_{12}\!\otimes\! M'_2\oplus L'_2\!\otimes\! M_2
$.
We select basis
$\{e^{12}_i\}\;\{g^{12}_{i'}\}$ in $L_{12}\;(M_{12})$ and complete
it with vectors
$\{e^1_j\}\;(\{g^1_{j'}\})$ to basis in $L_1\;(M_1)$ and with vectors
$\{e^2_m\}\;(\{g^2_{m'}\})$ to basis in $L_2\;(M_2)$. Note, that the system
$\{e_p\otimes g_q\}$,
where $e_p\in \{e^{12}_i\}\cup \{e^1_j\} \cup \{e^2_m\}$, and
$g_q\in \{g^{12}_{i'}\} \cup \{g^1_{j'}\} \cup \{g^2_{m'}\}$
is linear independent. So each of subspaces 
$A,B,C$ have zero intersection with the sum of two rest,
because they are linear closures of two by two non-intersecting subsystems 
of linear independent system. 
From this, in particular, follows that
$(A\!\oplus\! B) \cap(A\!\oplus\! C) = A$.  
Indeed, for all 
$x\in (A\!\oplus\! B) \cap (A\!\oplus\! C)$ 
we have $x=a+b=a'+c$ where
$a,a'\in A,\; b\in B,\; c\in C$, which implies $a-a'=c-b=0$, 
i.e.  $a=a',\; c=b=0$ and proof completed 
for case $s=2$. So, we proved equality:
\begin {equation} \label {f6}
(L_1\otimes M_1)\cap (L_2\otimes M_2) =
(L_1\cap L_2)\otimes (M_1\cap  M_2)
\end {equation}
Simple induction completes the proof.
\end{proof}

\begin {lemma} \label {lemma:l30}
	For an acyclic subspaces family $\{L_i\}_{i=\overline{1;s}}$
    the following formula is true:
	\begin {equation} \label{f7}
		\dim (L_1+\dots +L_s)=\sum_{\alpha\in
		S}(-1)^{|\alpha|+1}\dim L_{\alpha}
	\end {equation}
\end {lemma}
\begin{proof}
For $s\!=\!2$ our statement is the classic 
theorem on dimension of subspaces sum
(note that any pair of subspaces is acyclic family, which is wrong for triples).
Entire case can be received by simple induction, in respect that for acyclic
family the formula is true:
\begin {equation} \label {f10}
	( L_1 + \dots + L_{s-1} ) \cap L_s = ( L_1 \cap L_s ) + \dots + (L_{s-1}
	\cap L_s).
\end {equation}
Let make sure that the last formula is true. Indeed, inclusion
$
	( L_1 + \dots + L_{s-1} ) \cap L_s \supset ( L_1 \cap L_s ) + \dots +
	(L_{s-1} \cap L_s),
$
is, obviously, true without suggestion of acyclicty. In order to prove verse 
inclusion we choose a basis of family $\{L_i\}$ and consider decomposition of
arbitrary vector $y$ from subspace of left part of 
(\ref {f10}), by this basis. 
All of the basis vectors in this decomposition are belong to 
$L_s$, because basis is linear independent.
On other side by the same reason, each of these  vectors belongs to at least one
of subspaces $L_1,\dots L_{s-1}$, therefore,
$y$ belongs to subspace in right part of (\ref {f10}).
\end{proof}

\begin {lemma} \label {lemma:l40}
	For two acyclic families of subspaces $\B{C}$ и $\B{D}$,
	the family $\{C_i \otimes D_i\}$ is also acyclic.

\end {lemma}
\begin{proof}
Let $\{e_m\}$ be a basis in $\B{C}$ and $\{g_j\}$ be a basis
in $\B{D}$, then family of vectors $\{e_m \otimes g_j\}$ is 
linear independent, in view of families $\B{C}$ и $\B{D}$ are acyclic.
We choose the subfamily from this family by the following way:
for each $e_m \in C_{p_m}$ we take the products $e_m \otimes g_j$
with $g_j \in D_{p_m}$. It's easy to see, that resulting subfamily is the basis
of subspaces family $\{C_i \otimes D_i\}$. According to that the constructed family
is linear independent, it finishes the proof of acyclicty. $\;\triangleright$

Now it's easy to make sure that (\ref{f3}) is true.
Indeed, it's enough to apply lemma \ref{lemma:l30} to sum (\ref{f1}),
considering lemmas \ref{lemma:l20} and \ref{lemma:l40}.

For check inequality
(\ref{f4}) we simply enter the vector of weight $d_{\alpha}d'^{\alpha}$
or $d'_{\alpha}d^{\alpha}$, in code (\ref{f1}).

Let $x\in C_{\alpha}$ be a vector of minimal weight
$d_\alpha$ and $y\in D^\alpha$ be a vector of minimal weight
$d'^\alpha$. Then $y=y_{i_1}+\dots +y_{i_t}$, where
$t=|\alpha|$ and $y_{i_j}\in D_{i_j}$ and vector $x\!\otimes\! y = x\!\otimes\!
y_{i_1}+\dots+x\!\otimes\! y_{i_t}$ belongs to code (\ref{f1}) and have weight
$d_{\alpha}d'^{\alpha}$. The second case can be analyzed in the same manner.

And, finally, let us show that lower bound (\ref{f4g}) is true.
We consider an arbitrary vector
$x=\sum e_i \otimes b_i$ in $\B{C \otimes D}$ (see lemma \ref{lemma:l10}).  
Denote $\Psi_0 = \{\alpha\;|\; \alpha=\alpha_{e_i} , b_i \not=
0\}$. So far we consider standard matrix representation for a vector from
two spaces tensor product, exactly, if
$a=(a^1\dots a^n) $ и $c=(c^1\dots c^{n'}) $ is a pair of vectors from
$\B{F}^n$ and $\B{F}^{n'}$ respectively, then the coordinates of vector
$a\otimes c$ we write in matrix $(a_{ij})$, where $a_{ij}= a^i c^j$.
Then any vector $y$ in $\B{F}^{nn'} = \B{F}^n \otimes \B{F}^{n'}$
can be represented with sum of matrixes corresponding to it decomposable components.
The strings corresponding to nonzero positions of vector
$b = \cup {b_i}$ in matrix representation of $x$ are linear combinations
of vectors $e_i$. But this combinations are, obviously, belong to each
of subspaces $C^\beta$, where $\beta \in \Psi^*_0$ and, consequently, have the weight
greater or equal than $d^\beta$  for all $\beta$.
The weight of $b$ not less that $d'^{\alpha}$ for all $\alpha \in \Psi_0$,
because $b_i \in D^\alpha $ fore some $\alpha \in \Psi_0$.
The weight of vector $x$ is equal to sum of nonzero strings
of corresponding matrix. So we prove inequality:
$$
	\delta 	\geq
	\min_{\Psi_0 \subset \Psi(\B{e})}
		{\left[
			(\max_{\alpha \in\Psi_0} {d'^\alpha})
			(\max_{\beta\in\Psi^*_0} {d^\beta})
		 \right]
		}
$$

Vector $x$ can also be represented as $x=\sum a_i\otimes g_i$ 
and we can repeat above proof for this representation. 
Proof of theorem \ref{theorem:B} is completed.
\end{proof}

Let now family $\B{C}$ be acyclic and family $\B{D}$ be embedded. 
For an arbitrary vector $x \in \B{C \otimes D}$ let us check the verse inequality
for (\ref {f4}). First, we note that from lemma 
\ref {lemma:l11} it follows the existence of representation 
$ x = e_1\otimes b_1 + \dots + e_r\otimes b_r, $
for vector $x$ where $e_i$ is a in $\B{C}$ and $b_i\in D_{p_i}$, where
$p_i$ is maximal index with $e_i \in C_{p_i}$. 
Let us consider multi-index $\pi = \{p_i\} $. Let also $p = \min\pi$.
Then the weight of vector $b=\cup_i b_i$ not less than $d'_p$
(the distance of $D_p$ code). We remind, that $e_i \otimes b_i$ can be represented 
as matrix with strings, corresponding to nonzero positions of $b_i$, equal to $e_i$
and other strings are zeros. Respectively, vector $x$ can be represented as sum 
of such type matrixes and, consequently, the strings of $x$, corresponding to nonzero
components of $b$ are linear combinations of vectors $e_i$ and have weight not less than
$d^\pi$. So, the weight of vector $x$ not less than $d^\pi d'_\pi$.

Now let us show that lower bound (\ref {f4g}) and upper bound (\ref {f4}) are equal.
In our case, $\Psi(\B{g})=\{\{123\dots s\}, \{23\dots s\}, \dots, \{s\} \}$.
Let $\max_{\alpha\in\Psi_0}d^\alpha$ reached on some
$\alpha_0\in\Psi_0$ and say $\alpha_0=\{i_0,\dots,s\}$. 
Note that in this case we can consider
$\min_{\alpha\in\Psi_0}{|\alpha|} = |\alpha_0| $ and, consequently,
$i_0\in\alpha$ for all $\alpha\in\Psi_0$. 
From above we can conclude, that
$\beta_0=\{i_0\}\in\Psi_0^*$, thus
$\max_{\beta\in\Psi_0^*}d'^\beta\geq d'_{i_0} = d'_{\alpha_0}$
and we finally stay
\begin {displaymath}
	(\max_{\alpha\in\Psi_0} {d^\alpha})
	(\max_{\beta\in\Psi^*_0} {d'^\beta}) \geq d^{\alpha_0}d'_{\alpha_0}
\end {displaymath}
Proof of theorem \ref{theorem:B} is completed. \QED

The conclusion \ref{conc:c1} is a simple applying of theorems
\ref{theorem:A} and \ref{theorem:B} to the situation 
of two embedded code families.

\section {Some additions and examples.}
Here we consider some examples to see the behavior of upper and lower bounds.
Most of the examples are well-known (\ref{f1}).

The problem of calculating fractal code's distance in general case seems to be
very difficult. In our judgment the main problem related to fractal codes distance
is describing the obstacles of upper bound reaching.

\begin {example} \label {example:x1}{Golay code.} \end {example}

Let us consider the following code pairs:
\begin{displaymath}
\begin{array}{cccc}
C_1= & { \left( \begin{array}{llll}
	\verb ...11.11 \\
	\verb ..11.1.1 \\
	\verb .11.1..1 \\
	\verb 11.1...1 \\
	 \end{array} \right) }
	&
{C_2=} & {\left( \begin{array}{lll}
      \verb 1.11...1 \\
      \verb .1.11..1 \\
      \verb ..1.11.1 \\
      \verb ...1.111 \\
	\end{array}
	\right)
	}
	\\
D_1= & {\left(
	\begin{array}{l}
	\verb 111 \\
	\end{array}
	\right)}
	&
D_2=& {\left(
	\begin{array}{ll}
	\verb .11 \\
	\verb 11. \\
	\end{array}
	\right)}
\end{array}
\end{displaymath}
Where $C_1,\;C_2$ are (8,4,4)-codes, $D_1$ is (3,1,3)-code и $D_2$ is
(3,2,2)-code.

In this case $\B{C\otimes D}$ is (24,12,8)  Golay code with generating matrix:
\begin{displaymath}
	\B{C\otimes D}=\left({
			\setlength\arraycolsep{0pt}
			\begin{array}{llllllllllll}
		\verb ...11.11...11.11...11.11   \\
		\verb ..11.1.1..11.1.1..11.1.1   \\
		\verb .11.1..1.11.1..1.11.1..1   \\
		\verb 11.1...111.1...111.1...1   \\
		\verb ........1.11...11.11...1   \\
		\verb .........1.11..1.1.11..1   \\
		\verb ..........1.11.1..1.11.1   \\
		\verb ...........1.111...1.111   \\
		\verb 1.11...11.11...1........   \\
		\verb .1.11..1.1.11..1........   \\
		\verb ..1.11.1..1.11.1........   \\
		\verb ...1.111...1.111........   \\
			\end{array}
			}\right)
\end{displaymath}
The upper bound (\ref{f4}) is reached here.
Now we calculate lower bound (\ref{f4g}). We have 
$\Psi(\B{e})=\{1,2,12\},\quad\Psi(\B{g})=\{1,2\}$. Let us write the table
of possible values for $\Psi_0$ and $\Psi^*_0$ and corresponding values 
of inner maximums in (\ref{f4g}):
\begin {flushleft}
\begin {tabular} {|c|c|c|c|c|}
\hline
$\Psi_0\subset\Psi (\B {e}) $ 
	& $\{1,12\}$ 	& $\{2,12\}$ 	& $\{1,2\}$ 	& $\{1,2,12\}$\\
\hline
$ \Psi^*_0 $
	& $\{1,12\}$ 	& $\{2,12\}$ 	& $\{12\}$ 	& $\{12\}$\\
\hline
$ m_1 $
	& 12      	& 8 	      & 6   		& 6 \\
\hline
$\Psi_0\subset\Psi (\B {e}) $
	& $\{1\}$ 		& $\{2\}$ 		& $\{12\}$ 	& \\
\hline
$ \Psi^*_0 $
	& $\{1\}$ 		& $\{2\}$ 		& $\{1,2\}$	& \\
\hline
$ m_1 $
	& 12      		& 8       		& $\B{4}$	& \\
\hline
\hline
$\Psi_0\subset\Psi (\B {g}) $
	& $\{1\}$ 		& $\{2\}$       & $\{1,2\}$		& \\
\hline
$ \Psi^*_0 $
	& $\{1\}$ 		& $\{2\}$       & $\{12\}$		& \\
\hline
$ m_2 $
	& 12			& 8        		& $\B{4}$	& \\
\hline

\end {tabular} \\
где $ m_1 =	(\max_{\alpha \in\Psi_0} {d'^\alpha})
			(\max_{\beta\in\Psi^*_0} {d^\beta})
	$ и   \\
	$ m_2 = (\max_{\alpha \in\Psi_0} {d^\alpha })
			(\max_{\beta\in\Psi^*_0} {d'^\beta }).
	$
  The values of minimums in (\ref{f4g}) are bolded.
\end {flushleft}

From the above table we see that lower bound in 
this example is equal to $\max{(4,4)}=4$.

\begin {example}\label{example:x2}
	(21,12,5)-code
\end {example}

Excluding last column in codes $C_1 $ and $C_2$ of previous example
we get (7,4,3)-codes with intersection containing one nonzero vector of weight 7.
If not changing codes $D_1$ and $D_2$, we get (21,12,5)-code as
$\B{C\otimes D}$. In this case bound (\ref{f4}) has value of 6 and not reached.
But received code is optimal (we can't increase $d$).
Lower bound is equal to $\max{(3,4)}=4$ in this case.

\begin {example} \label {example:x3}
	(21,8,9)-code
\end {example}

Excluding next to the last column and last row in code $C_2$ 
from example \ref{example:x1} and 
next to the last column and first row,
in code $C_1$ from example \ref{example:x1}, we get the pair of 
(7,3,4)-codes with zero intersection. The corresponding 
$\B{C\otimes D}$ codes has the parameters (21,9,8).  The bound (\ref{f4})
is reached and we have optimal code again. The lower distance bound
(\ref {f4g}) is equal to $\max{(4,6)}=6$ in this case.

\begin {example} \label {example:x4}
	(28,22,4)-code
\end {example}

Let us take (7,3,4)-code $C_2$ from previous example as $C_1$, the (7,6,2)-code
of all even-weight vectors as $C_2$ and (7,7,1)-code of all vectors as $C_3$.
Note that received codes' family is embedded. 

Then denote (4,4,1)-code of all vectors as $D_1$, 
(4,3,2)-code of all even-weight vectors as $D_2$ and uniquely defined
(4,1,4)-code as $D_3$. According to theorem \ref{theorem:B}, 
$\B{C\otimes D}$ is (28,22,4)-code.
The upper bound (\ref{f4}) is reached and equal to lower bound.
This code is optimal.

Let us make direct calculation of lower bound by formula (\ref{f4g}).
In this case we have
$\Psi(\B{e})=\{123,23,3\},\quad\Psi(\B{g})=\{1,12,123\}$.
Because of situation is symmetric in relation to
$\B{e}$ and $\B{g}$, it is enough to consider only the first case. 
There is the table for this case (see analog table in example \ref{example:x1}):
{ \footnotesize
\begin {center}
\begin {tabular} {|c|c|c|c|c|}
\hline
$\Psi_0\subset\Psi (\B {e}) $
& $\{123\}$ 				& $\{23\}$ 			& $\{3\}$
& $\{123,\B{23}\}$          \\
\hline
$\Psi^*_0 $
& $\{1,2,3\}$				& $\{2,3\}$ 		& $\{3\}$
& $\{12,13,\B2,23,\B3\}$    \\
\hline
$ m_1 $
	& 4      		& 4       		& 4
	& 4      		\\
\hline
$\Psi_0\subset\Psi (\B {e}) $
& $\{123,\B 3\}$ 	& $\{23,\B 3\}$
& $\{123,23,\B3\}$	&	\\
\hline
$\Psi^*_0 $
& $\{13,23,\B3\}$ 		& $\{23,\B3\}$
& $\{123,13,23,\B3\}$	&	\\
\hline
$ m_1 $
& 4       		& 4   			& 4  & \\
\hline
\end {tabular}
\end {center}
}

Note that it is enough to consider only minimal by inclusion indexes. They
are bolded in the table. So, we made sure in equality of upper and lower bounds
by direct calculation.

\begin {example}
	(32, 16, 8)-code.
\end {example}

Let us consider two embedded codes families
$C_1 \subset C_2 \subset\ C_3 $ and
$D_1 \subset D_2 \subset\ D_3 $ with following parameters:

\begin {tabular} {|c|c||c|c|}
	\hline
	код & $(n,k,d)$ & код & $(n,k,d)$ \\
	\hline
	$C_1$ & (4,1,4) & $D_1$ & (8,1,8) \\
	\hline
	$C_2$ & (4,3,2) & $D_2$ & (8,4,4) \\
	\hline
	$C_3$ & (4,4,1) & $D_3$ & (8,7,2) \\
	\hline
\end {tabular}

All codes are uniquely defined, with exception of $D_2$ code, which can be
choose arbitrarily. According to conclusion \ref{conc:c1}, 
the code defined by formula 
(\ref{f6}) is $(32,16,8)$-code. This code is equivalent to 
Reed-Muller code $\mathcal{R}(5,2)$ and is optimal.

\begin {example} \label{example:x5}
	$|u|u+v|$ construction (see. ~\cite{mac} \S 2.9)
\end {example}

Let $C_1$ and $C_2$ be arbitrary codes, $D_1$ be obvious
(2,1,2)-code and $D_2$ be (2,1,1)-code containing vectors (0, ~0) and (0, ~1).
Then $\B{C\otimes D}$ is $|u|u+v|$ construction. In this case,
upper and lower bounds are equal to $\min{(2d_1, d_2)}$, where
$d_1,\;d_2$ are $C_1,\;C_2$ distances respectively.

\begin {example} \label {example:x6}
	$|a+x|b+x|a+b+x|$ construction (see ~\cite{mac} \S 18.7.4)
\end {example}

Again $C_1$ and $C_2$ are arbitrary codes. $D_1$ and $D_2$ are 
uniquely defined (3,1,3)- and (3,2,2)-codes respectively. 
Then $\B{C\otimes D}$ is $|a+x|b+x|a+b+x|$ construction.
The upper bound can not always be reached in this case
(see previous examples). For the first time the Golay's code construction
using $|a+x|b+x|a+b+x|$ was given by Turyn ~\cite{tur}.
Lower bound in this case is depended on $\{C_i\}$ family configuration.

The following two problems related to fractal codes seems to be interesting:
determining of entire conditions of upper bound reaching and determining necessary
and enough conditions of code to be equal (or equivalent) to fractal code. The second 
problem in particular case of tensor product code was formulated and solved in \cite{bar}.  

We would like to put on record our indebtedness to academician H.~H.~Khachatryan
from whom we learnt the subject, and whose influence was the
determining factor in our choice of error correcting codes theory as research subject.

\end {document}